\documentclass[12pt]{article}
\usepackage{epsfig}

\def\beq{\begin{equation}}
\def\eeq#1{\label{#1}\end{equation}}
\def\eeqn{\end{equation}}

\def\beqa{\begin{eqnarray}}
\def\eeqa#1{\label{#1}\end{eqnarray}}
\def\eeqan{\end{eqnarray}}

\let\bar=\overbar

\def\Dslash{\not{\hbox{\kern-4pt $D$}}}
\def\dslash{\not{\hbox{\kern-2pt $\del$}}}

\def\msb{{\bar{\ssstyle M \kern -1pt S}}}

\def\Title#1{\begin{center} {\Large {\bf #1} } \end{center}}

\newcommand{\msun}{$M_{\odot}\ $}

\begin{document}

\Title{Structure of rapidly rotating strange stars: salient differences from
neutron stars}

\bigskip\bigskip

%+\addtocontents{toc}{{\it D. Reggiano}}
%+\label{ReggianoStart}

\begin{raggedright}  

{\it Arun V. Thampan
\index{Thampan, A.V.}\\
SISSA, \\
via Beirut n.2--4, \\
Trieste 34014, \\
ITALY}
\bigskip\bigskip
\end{raggedright}

\section{Introduction}

The existence of a third class of compact objects (apart from white dwarfs and
neutron stars) has been debated for some time now (e.g. [1]).  
Attention may be drawn, particularly, to suggestions of ``bare'' strange 
stars as the candidates [2].  Although an 
important question in this regard would be the
formation mechanisms for such objects, an issue 
of equal import is that of observationally distinguishing these 
objects from neutron stars and black holes.  In this work, we
juxtapose the parameters of structure of compact stars (``bare''
strange stars and neutron stars) in an attempt to glean the salient 
differences between these two classes of compact object  - a first step 
in proving (or disproving) the existence of strange stars.  

The discussion presented here is, therefore, expected to have relevance when 
modelling future sensitive observational data from low mass X--ray binaries
(LMXBs). LMXBs are compact objects in binary orbit with an evolved
low mass or dwarf companion star.  The closeness of the orbit
permits the compact object to peel off the outer layer of the
companion star.  Owing to it possessing substantial angular momentum, the 
matter so accreted (or peeled off), forms an accretion disk around the
compact star.

Astrophysical models of LMXBs (in particular, the X--ray burst sources) play 
an important role in the constraining the equation of state of high density 
matter of the central object.  It is, therefore, imperative that the models 
so constructed be realistic.  Including general relativisitic 
effects is inescapable.  General relativity not only decides the internal 
structure of compact stars, but also their external spacetime.

\begin{figure}[htb]
\begin{center}
\vspace{-1.5 in}
\epsfig{file=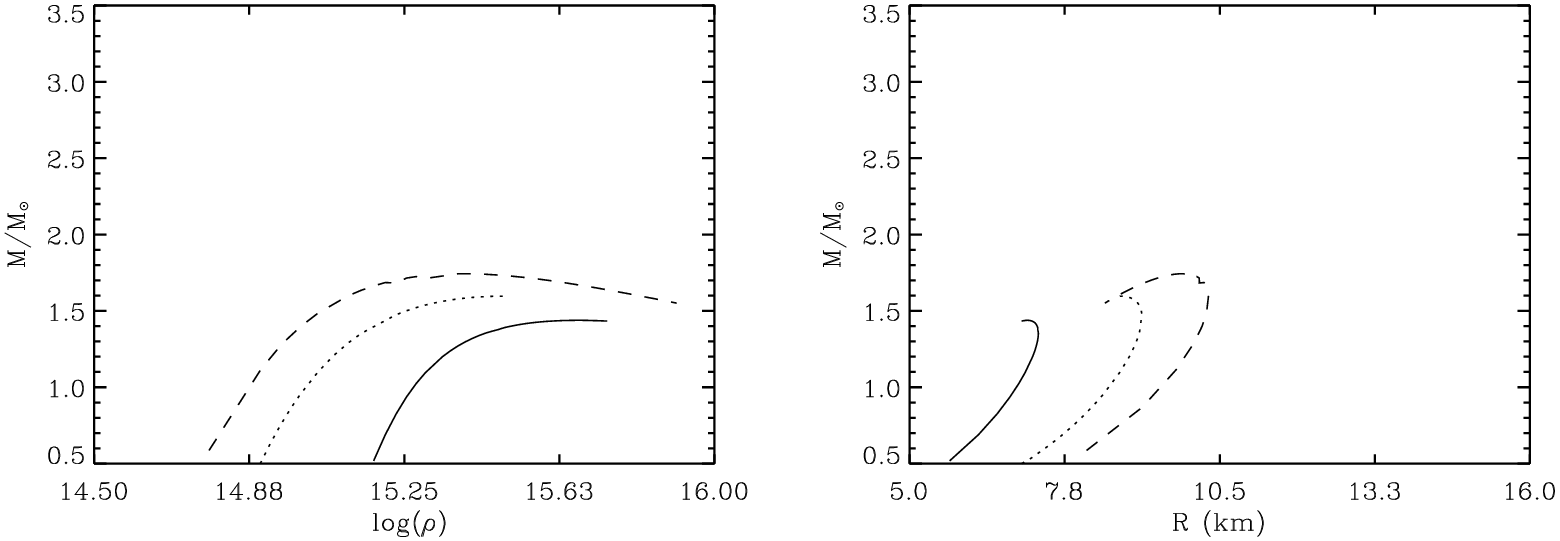,height=5.0in}\\
\vspace{-1.8 in} \hspace{1.in} (a)\\
\vspace{-1.8 in} %\hspace{+0.25 in}
\epsfig{file=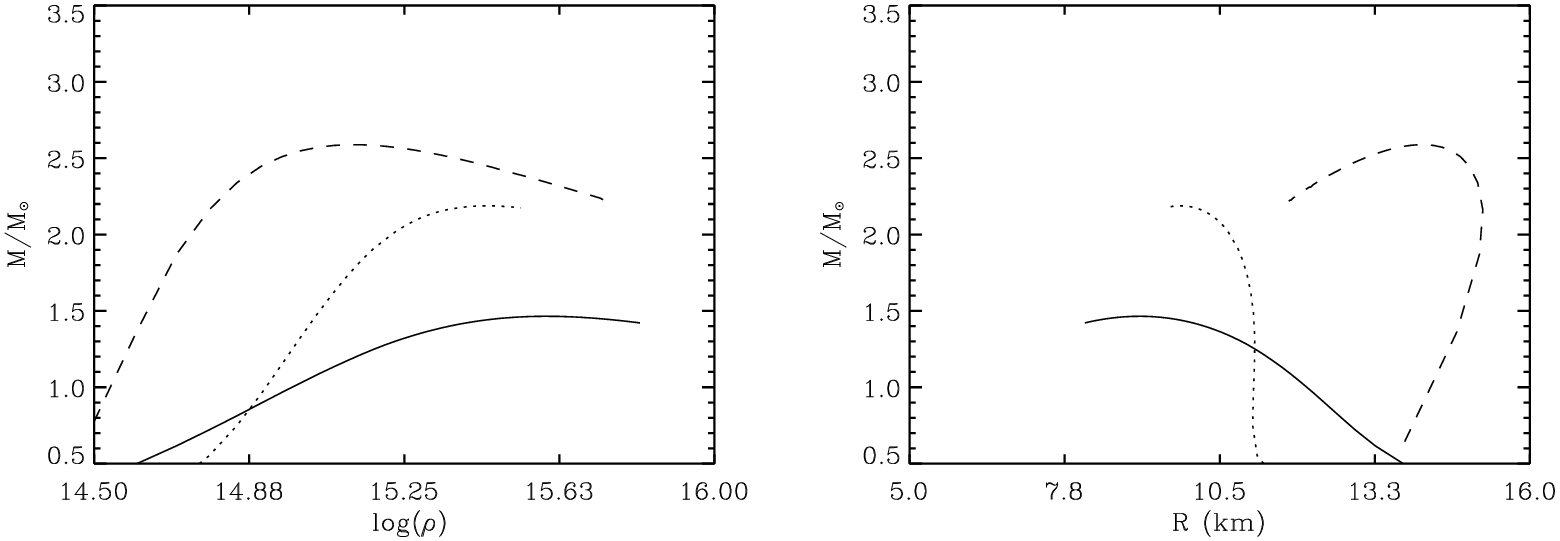,height=5.0in}\\
\vspace{-1.8 in} \hspace{1.in} (b)\\
\caption{Structure parameters for non--rotating compact stars.  The
upper panel corresponds to strange stars and the lower to neutron
stars. The curves correspond to: Upper Panel: EOS A (solid), B
(dotted), C (dashed) and Lower Panel: EOS BPAL12 (solid), WFFII
(dotted) and SBD (dashed) respectively.}
\label{fig:ssstruc}
\end{center}
\end{figure}

General relativity implies the existence of an innermost marginally
stable orbit around compact objects with a radius $r_{\rm in}$.  For 
non--rotating (Schwarzschild) compact objects, $r_{\rm in}=3GM/c^2$.  
For rotating compact objects, the $r_{\rm in}$ decreases (for a given 
rest mass) with increasing angular momentum.  Accretion disks around
compact objects will have an inner--edge located at $r=r_{\rm in}$.  
Neutron stars as well as strange stars may have radii $R>r_{\rm orb}$.  

\begin{figure}[htb]
\begin{center}
%\vspace{-1.0 in}
\epsfig{file=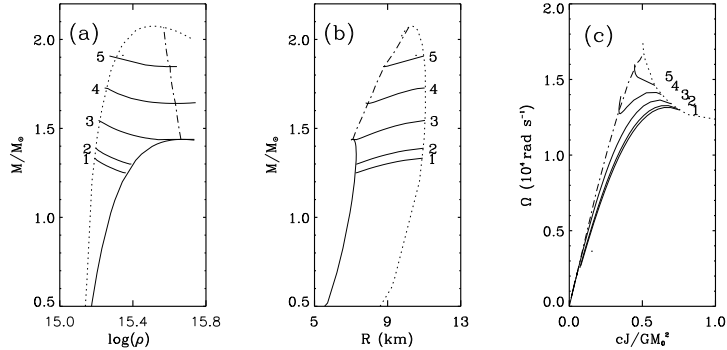,height=2.0in}
\caption{Rapidly rotating strange star structure parameters. EOS model
used is EOS A. Labels 1--5 correspond to $M_0= 1.59$, $1.66$, $1.88$,
$2.14$, $2.41$ (where $M_0$ is the baryonic mass and the numbers are
in units of solar mass \msun: $1.9\times10^33$ gms) respectively.}
\label{fig:rots}
\end{center}
\end{figure}

The matter spiralling in from the inner--edge of the accretion disk, 
transfers angular momentum to the compact star, spinning it up to high 
rotation rates ($\sim$ millisecond periods) over dynamical timescales.  
Realistic astrophysical models must incorporate the effects of rotation 
of the accreting compact object.

\begin{figure}[htb]
\begin{center}
%\vspace{-1.5 in}
\epsfig{file=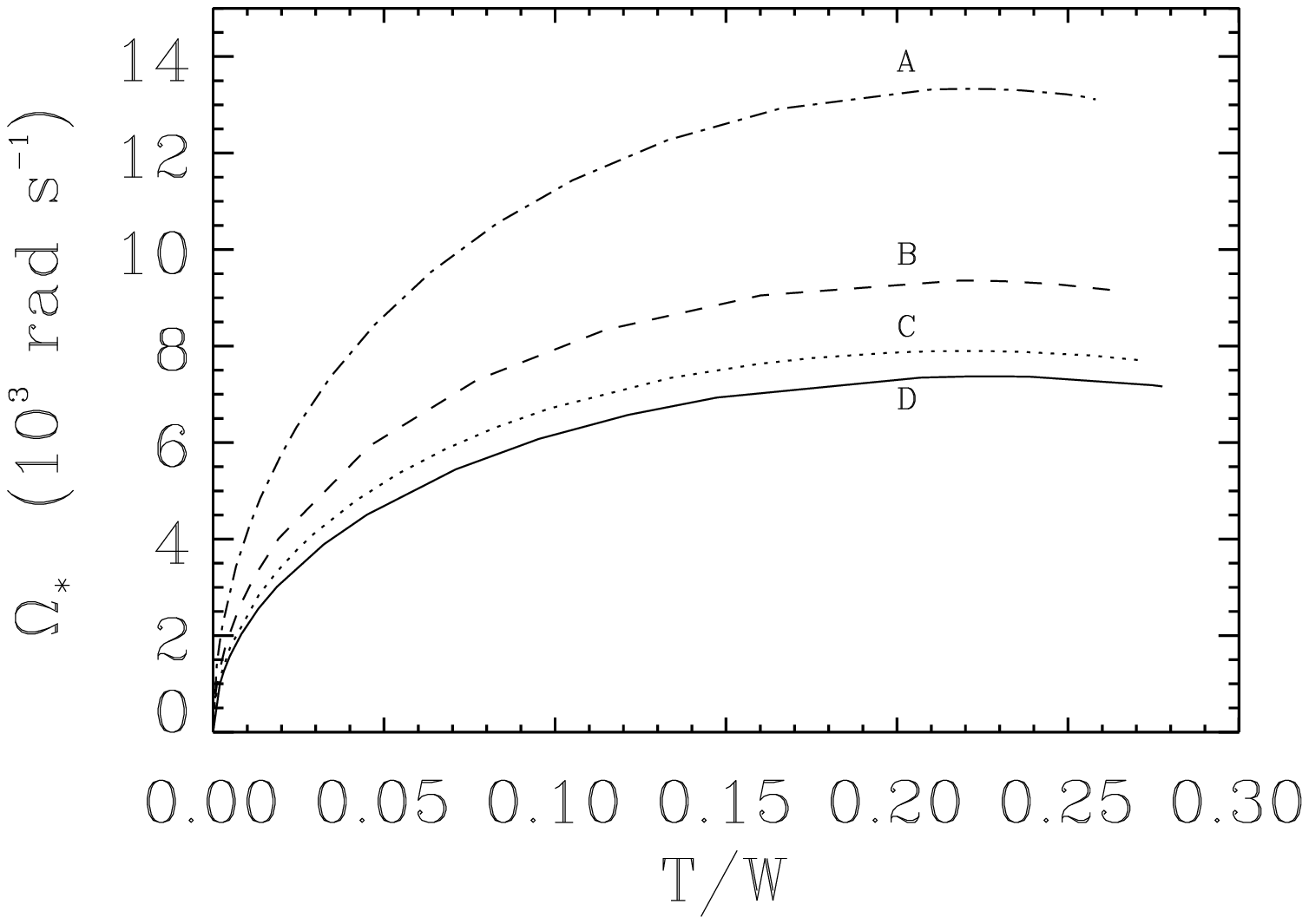,height=2.0in}
\caption{Strange star rotation rate as a function of the ratio of
rotational kinetic energy to the total gravitational energy.  The
labels on curves correspond to the strange star EOS models used. 
All curves correspond to gravitational mass sequence $M=1.4$ \msun.} 
\label{fig:omtbw}
\end{center}
\end{figure}

In this work, we point out some of the salient differences between 
rotating neutron and strange stars.  We refer the reader to  [3], [4],
[5], for details on the results presented here.

\begin{figure}[htb]
\begin{center}
%\vspace{-1.5 in}
\epsfig{file=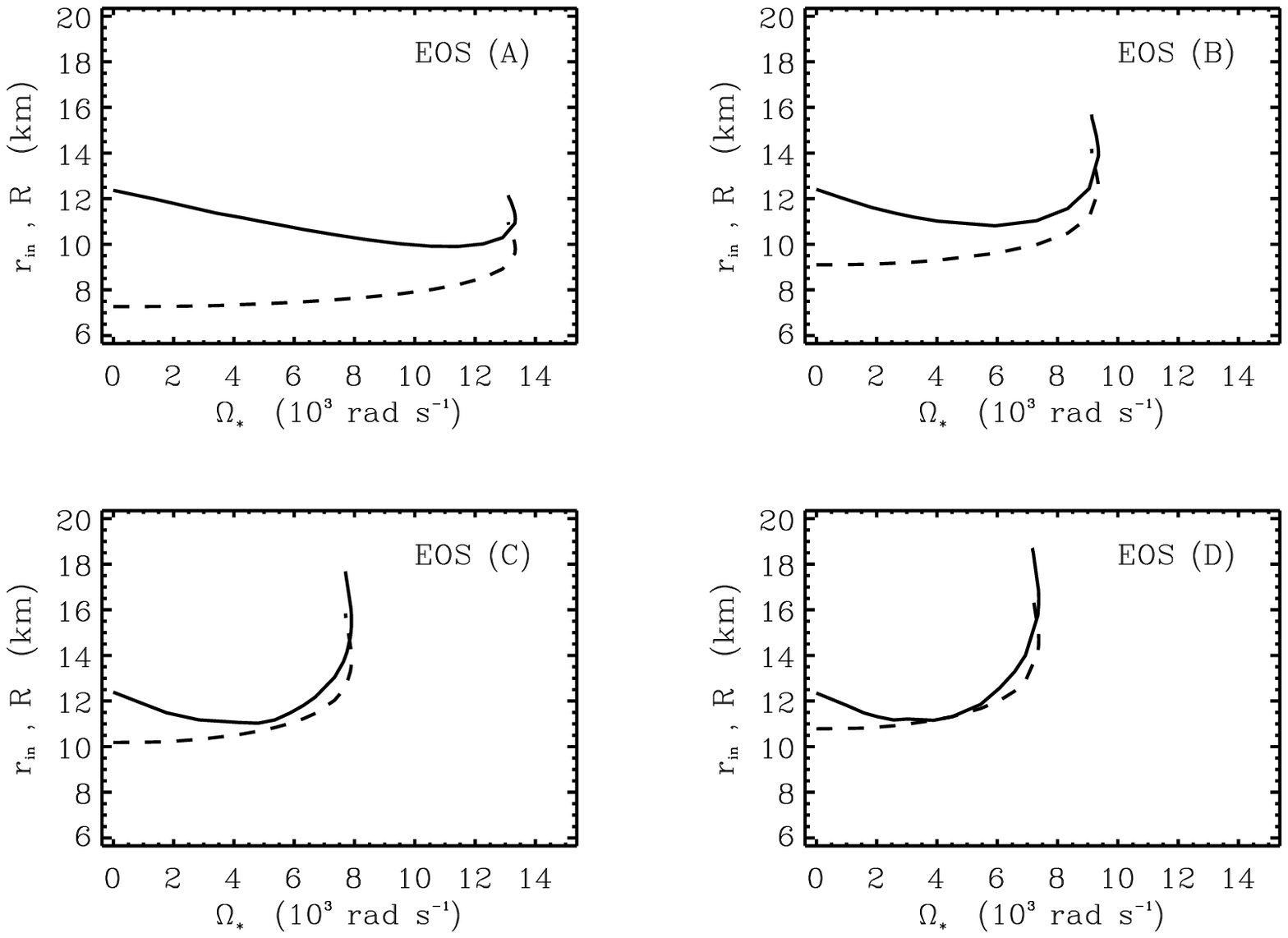,height=2.0in}
\caption{Strange star radii ($R$) and innermost stable circular 
orbit radii ($r_{\rm in}$) variations with rotation rate for
graviational mass sequences corresponding to $M=1.4$ \msun.} 
\label{fig:Rrinom}
\end{center}
\end{figure}

\section{Rotating compact stars}

\begin{figure}[htb]
\begin{center}
%\vspace{-1.5 in}
\epsfig{file=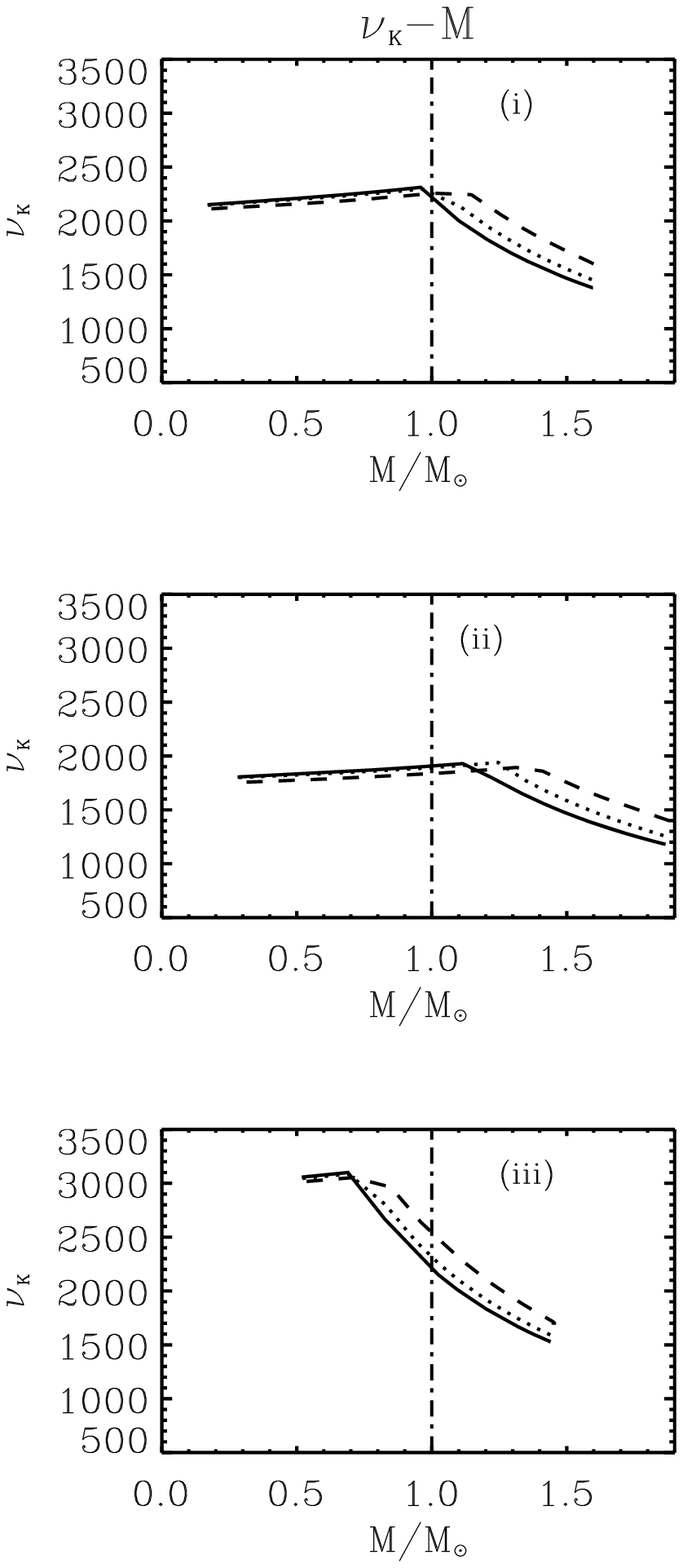,height=3.0in}
\caption{Kepler frequencies of test particles around rotating strange
stars. The panel (i) corresponds to EOS B, and panel (iii) to A.
Panel (ii) corresponds to EOS model due to [11].  The curves in 
each of these panel correspond to strange star rotation rates of
0 (solid), 200 (dotted) and 580 (dashed) Hz. The vertial dot--dashed
lines in the panel represents a 1 \msun configuration.}
\label{fig:nuk}
\end{center}
\end{figure}

The first calculation of the structure of non--rotating bare strange stars 
[2], established the mass-radius
relationship to be the chief difference between the structure of
non--rotating strange stars and neutron stars (Fig. 1).  
In the calculations reported here, we make use of 4 equations of state (EOS) 
models for strange stars (in what follows $B$, $\alpha_{\rm c}$, $m_{\rm s}$,
$m_{\rm u}$, $m_{\rm d}$ refer to the Bag constant, QCD structure
constant, and masses of strange, up and down quarks respectively): 
EOS A: [6], and MIT Bag model [7] 
with EOS B: $B= 90$ MeV/fm$^3$, EOS C: $B = 60$ MeV/fm$^3$, 
$\alpha_{\rm c} =0$, $m_{\rm s} = 200$~MeV and $m_{\rm u} = m_{\rm d} = 0$,
EOS D: $B = 60$ MeV/fm$^3$, $\alpha_{\rm c} =0$
and 3 EOS models for neutron stars: BPAL12 [8];
WFFII [9]; SBD [10].
NSEOS model SBD is a very stiff EOS, while WFFII is moderate in stiffness and
BPAL12 is very soft.  In addition to these EOS models, we also use
(for displaying the effect of EOS on Kepler frequencies) a strange
star EOS model due to [11].

To construct rapidly rotating strange stars, we perform numerical
computations based on the formalism by [12].  We construct sequences 
all the way from the static
limit to mass--shed limit where centrifugal forces balance the inward
gravitational pull.  We also construct constant gravitational mass
($M$) and constant baryonic mass ($M_{\rm B}$) sequences.  The results 
of our calculations, for one SSEOS: ... is shown in Fig. 2: the thick 
solid curve is the non--rotating limit, the dotted curve is the mass 
shed limit and the horizontally directed curved lines are constant 
baryonic mass sequences.

The constant gravitational mass sequences help studying the effect of
spin on the strange stars.  From Fig. 3, we notice that 
there exists a maximum rotation rate for the strange stars, with a 
value that is higher than the rotation rate at mass shed (an 
effect that is absent in neutron stars: [13]).  
This maximum occurs at $T/W \sim 0.2$, indicating that such angular 
velocities are perhaps not attainable in strange stars due to triaxial 
instabilies [14].

Fig. 4 showing $r_{\rm in}$ v/s $\Omega$ for constant gravitational mass 
sequences imply that  $r_{\rm in}$ decreases with increasing 
$\Omega_{*}$ for low values of the rotation rate (similar to Kerr black 
holes and rotating neutron stars), but reaches a minimum and increases 
for higher rotation rates.  The increase in $r_{\rm in}$ with rotation 
is purely an effect of the large quadrupole moment of the strange star 
[15]. 

We also compute the Kepler frequencies of test particles around
rotating strange stars (Fig. 5).  On comparing these with the Kepler 
frequencies expected for neutron stars (see [16]), we see that 
frequencies in the range 
(1.9 - 3.1) kHz if observed in future in X--ray burst sources, 
can be understood in terms of strange stars rather than neutron
stars. 

\noindent Acknowledgements:  The results presented in this paper was 
obtained at Inter--University Centre for Astronomy and Astrophysics 
(IUCAA), Pune, India.

\def\Discussion{
\setlength{\parskip}{0.3cm}\setlength{\parindent}{0.0cm}
     \bigskip\bigskip      {\Large {\bf Discussion}} \bigskip}
\def\speaker#1{{\bf #1:}\ }
\def\endDiscussion{}

\Discussion

\endDiscussion
 
\end{document}